\documentstyle[multicol,prl,aps,psfig,floats]{revtex}
\begin{document}
\tightenlines
\title{Coulomb Interaction and Quantum Transport through a Coherent Scatterer}
\author{Dmitrii S. Golubev and Andrei D. Zaikin}
\address{Forschungszentrum Karlsruhe, Institut f\"ur Nanotechnologie,
76021 Karlsruhe, Germany\\
I.E.Tamm Department of Theoretical Physics, P.N.Lebedev
Physics Institute, Leninskii pr. 53, 117924 Moscow, Russia}

\maketitle

\begin{abstract}
An interplay between charge discreteness, coherent scattering and Coulomb
interaction yields nontrivial effects in quantum transport. We derive a
real time effective action and an equivalent quantum Langevin equation for
an arbitrary coherent scatterer and  evaluate its current-voltage characteristics
in the presence of interactions. Within our model, at 
large conductances $G_0$ and low $T$ (but outside the 
instanton-dominated 
regime) the interaction correction to $G_0$ saturates and causes 
conductance suppression by a universal factor 
which depends only on the type of the conductor.

\end{abstract}


\begin{multicols}{2}


Coulomb effects in mesoscopic tunnel junctions have recently received 
a great deal of attention \cite{AL,SZ,GD,so97}. One of the remarkable features 
of such systems is that charge quantization (and, hence, Coulomb blockade) 
persists even for junctions with low resistances 
$1/G_t \ll R_{Q}=h/e^2\approx 25.8$ k$\Omega$. In this limit 
an effective Coulomb gap $\tilde E_C$ for a junction with the ``bare'' 
charging energy $E_C$ suffers exponential renormalization \cite{pa91}
\begin{equation}
\tilde E_C/ E_C \propto \exp (-G_tR_Q/2),
\label{rengap}
\end{equation}  
but remains finite even at very large values of $G_tR_Q$. 
Eq. (\ref{rengap}) was confirmed
in several later studies both analytically \cite{WG,ho97} and numerically \cite{ho97,ca99}. 
Experiments 
clearly demonstrated the existence of charging effects for the values of $G_t$
as large as $G_tR_Q \approx 32$ \cite{Chalmers}.

Recently another interesting prediction was made by Nazarov \cite{Naz},
who argued that features of charge quantization 
may also persist in arbitrary conductors including, e.g., 
disordered metallic wires 
with $g=G_0R_Q \gg 1$. Here and below 
$G_0\equiv 1/R=(2e^2/h)\sum_nT_n$ is the conductance of an arbitrary scatterer
and  $T_n$ are the transmissions
of its conducting modes. If one accounts for the spin degeneracy, 
the renormalized Coulomb energy for a general conductor derived in 
\cite{Naz} takes the form 
\begin{equation}
\tilde E_C/E_C \propto \prod_{n}R_n,
\label{rengap2}
\end{equation}    
where $R_n=1-T_n$. In particular, 
for diffusive conductors, similarly
to eq. (\ref{rengap}), one finds \cite{Naz} $\tilde E_C/E_C\propto \exp (-\pi^2g/8)$. 
The same result (\ref{rengap2}) follows from 
the effective action derived in \cite{SZ,Z} for metallic contacts within
the quasiclassical Green functions  technique. Hence, one can expect the
effective actions \cite{SZ,Z}  and \cite{Naz} to be equivalent, perhaps up to
some unimportant details.  

Eq. (\ref{rengap2}) sets an important energy scale for the problem in question:
at temperatures below an exponentially small value $\tilde E_C$ a conductor 
with $g \gg 1$ should show {\it insulating} behavior due to Coulomb effects.
On the other hand, at larger temperatures/voltages this insulating
behavior should not be pronounced. Furthermore, according to
(\ref{rengap2}) Coulomb blockade is destroyed completely ($\tilde
E_C\equiv 0$) even at $T=0$ if at least one of the conducting channels is fully
transparent $R_n=0$ \cite{M}.

In this Letter we will analyze an interplay between Coulomb effects and quantum
transport at energies larger than $\tilde E_C$ (\ref{rengap2}).
We will derive a
real time effective action and formulate a quantum Langevin equation
for an arbitrary (albeit relatively short) conductor. At temperatures or
voltages above $\tilde E_C$  we will obtain a complete $I-V$ curve 
at large enough $g$. We will demonstrate that Coulomb
interaction leads to (partial) conductance suppression with respect
to its ``noninteracting'' value $G_0$. This suppression effect is 
controlled by the parameter 
\begin{equation}
\beta=\frac{\sum_nT_n(1-T_n)}{\sum_nT_n},
\label{param}
\end{equation}
well known in the theory of shot noise \cite{BB}. 
The parameter $\beta$ (\ref{param}) equals to one for tunnel
junctions and to $1/3$ for diffusive conductors. In contrast to 
$\tilde E_C$ (\ref{rengap2}), it vanishes only if {\it all} the conducting 
channels are fully transparent.

We identify three different regimes for the interaction correction to $G_0$.
Let us display the results for
a linear conductance $G(T)$.
At $T/E_C \gg$ max$(1,g)$ perturbation theory in $E_C$ (or in $1/T$) is
sufficient.  It yields
\begin{equation} 
\frac{G}{G_0}\simeq
1-\beta\left\{\frac{E_C}{3T}-\left(\frac{3\zeta(3)}{2\pi^4}g+ 
\frac{1}{15} \right)\left(\frac{E_C}{T}\right)^2\right\}.
\label{largeT}
\end{equation}
Here $\zeta(3)\simeq 1.202$ and $g$ needs not to be necessarily large. For $g
\gg 1$ there exist two further nonperturbative in the interaction regimes. At
intermediate temperatures $gE_C\exp(-g/2)\ll T \ll gE_C$ we have
\begin{equation} 
\frac{G}{G_0}\simeq 1-\frac{2\beta}{g}\left[\gamma
+1+\ln\left(\frac{gE_C}{2\pi^2 T} \right) \right], 
\label{intT}
\end{equation}
where $\gamma \simeq 0.577$. Here energy relaxation plays an
important role turning the power law dependence (\ref{largeT}) into a much
slower one (\ref{intT}). Finally, at even lower temperatures $T <
gE_C\exp(-g/2)$ (but $T>\tilde E_C$) relaxation processes yield complete
saturation of $G(T)$: 
\begin{equation} 
G/G_0\simeq 1-\beta +O(\beta /g).
\label{lowT}
\end{equation}
It is remarkable that the result (\ref{lowT}) does not depend on
the charging energy $E_C$ at all. In the tunneling limit (all $T_n \ll 1$) 
the regime (\ref{lowT}) does not exist. Two other regimes are already 
known for tunnel junctions: 
by setting $\beta =1$ in eqs. (\ref{largeT}), (\ref{intT}) we recover 
the results \cite{GZ96,KSS,GG}.

{\it The model and effective action}. Now let us proceed with the derivation
of the above results and the $I-V$ curve. Our framework is very similar 
to that of Ref.
\cite{Naz}. We will consider an arbitrary scatterer between two big reservoirs.
The scatterer length is assumed to be shorter than dephasing and inelastic
relaxation lengths, so that phase and energy relaxation may occur only in the
reservoirs and not during scattering. Coulomb effects in
the scatterer region are described by an effective
capacitance $C$. The charging energy $E_C=e^2/2C$, temperature $T$ 
as well as other energy scales are assumed to
be smaller than the typical inverse scattering time (e.g. the Thouless energy
in the case of diffusive conductors).

Quantum dynamics of our system is fully described by the evolution operator
on the Keldysh contour. The kernel of this operator $J$ may be represented
as a path integral over the fermionic fields. Performing a standard
Hubbard-Stratonovich decoupling of the interacting term in the Hamiltonian
enables one to integrate out fermions. Then the kernel $J$ acquires the form
of the path integral over the Hubbard-Stratonovich fields on the forward
($V_1$) and backward ($V_2$) parts of the Keldysh contour  
\begin{equation}
J=\int {\cal D} V_1{\cal D}V_2\exp(iS[V]),
\label{pathint}
\end{equation}
where $S[V]$ is the effective action defined as
\begin{eqnarray}
iS[V]=2{\rm Tr}\ln \widehat{\bf G}^{-1}_V 
+i \frac{C}{2}\int\limits_0^t dt'  (V_{LR1}^2-V_{LR2}^2),
\label{S}
\end{eqnarray}
where $V_{LRi}\equiv V_{Li}-V_{Ri}$ are the voltage drops
between the reservoirs. The Green-Keldysh matrix $\hat{\bf G}_V(X_1,X_2)$ 
(here $X=(t,\bbox{r})$) obeys the 
2$\times$2 matrix equation
\begin{equation}
\left[i\frac{\partial}{\partial t_1}\hat{\bf 1}-\widehat
H_0(\bbox{r}_1)\hat{\bf 1}+e\widehat {\bf V}(X_1)\right] \widehat{\bf G}_V=
\delta(X_1-X_2)\hat{\bbox{\sigma}}_z,
\label{Schrodinger}
\end{equation}
where $\widehat H_0(\bbox{r})$ is a free electron Hamiltonian for the system 
``scatterer + reservoirs'', $\widehat{\bf V}$ is the diagonal 2$\times$2 matrix with
the elements ${\bf V}_{ij}=V_i\delta_{ij}$ and $\hat{\bbox\sigma}_z$ is the Pauli matrix. In the last term of eq. 
(\ref{S}) we already made
use of our model and assumed that the fields $V_{1,2}$ do not depend on the coordinates
inside the reservoirs, i.e. for the left (right) reservoir we put
$V_{j}(t',\bbox{r})\equiv V_{L(R)j}(t')$.

The elements of the Green-Keldysh matrix $\widehat{\bf G}_V$ can be expressed as follows:
\begin{eqnarray}
\widehat G_{11}(t_1,t_2)&=&-i\theta(t_1-t_2)\widehat U_1(t_1,t_2)+i\widehat U_1(t_1,t)\widehat\rho(t)\widehat U_1(t,t_2),
\nonumber\\
\widehat G_{21}(t_1,t_2)&=&-i\widehat U_2(t_1,t)(1-\widehat\rho(t))\widehat U_1(t,t_2),
\label{Gij}
\end{eqnarray}
and similarly for $\widehat G_{12}$ and $\widehat G_{22}$. 
Here and below integration over the spatial coordinates is
implied in the products of operators.
In (\ref{Gij}) we have defined  
\begin{equation}
\widehat U_{j}(t_1,t_2)=\hat T \exp\left[-i\int\limits_{t_1}^{t_2}
dt'\left(\widehat H_0 -eV_{j}(t',\bbox{r})\right)\right],
\label{ev}
\end{equation}
as the evolution
operators \cite{GZ99} and $\widehat\rho(t)$ as the density matrix. The latter satisfies an exact equation \cite{GZ99}
\begin{equation}
i\frac{\partial}{\partial t}\widehat\rho = [\widehat H_0,\widehat\rho]-(1-\widehat\rho)eV_1\widehat\rho
+\widehat\rho eV_2(1-\widehat\rho),
\label{eqrho}
\end{equation}
with the initial condition $\widehat\rho(t=0)=\widehat\rho_0,$ where $\widehat\rho_0$ is the equilibrium density matrix for noninteracting electrons. 

Next we define the conducting channels in a standard manner. They are just 
the transverse quantization modes in the reservoirs.
Describing the longitudinal motion within one channel
quasiclassically we define the free electron Hamiltonian
in the reservoirs as follows
\begin{equation}
\widehat H_{0, m n } = -i v_{m}\delta_{m n }\frac{\partial}{\partial y},
\label{ham}
\end{equation}
where $m ,n $ are the channel indices
and  $v_{m}$ is the channel velocity. In every channel the coordinate $y$ runs
from $-\infty$ to $0$ for the incoming waves, and from $0$ to
$+\infty$ for the outgoing ones. The scattering
matrix $\widehat S$, which is assumed here to be energy independent, 
relates the amplitudes of incoming and outgoing modes as follows
\begin{equation}
\psi_m(y=+0)=\sum\limits_n S_{mn}\sqrt{\frac{v_n}{v_m}}\, 
\psi_n(y=-0),
\label{bound}
\end{equation} 
where $S_{mn}$ are the elements of the scattering matrix $\widehat S$
defined in the basis $\psi_{0,m}={\rm e}^{iky}/\sqrt{v_m}$.
The factor $\sqrt{{v_n/v_m}}$ appears in (\ref{bound}) since
we work in the basis of the eigenfunctions of (\ref{ham}) $\psi_{0,m}={\rm
e}^{iky}$.  Finally, the matrix elements of the fluctuating voltages $V_j(t)$
are: $V_{j,mn}(t)=V_{j,m}(t)\delta_{mn},$ 
where $V_{j,m}(t)=V_{Lj}(t)$ for the left channels and
$V_{j,m}(t)=V_{Rj}(t)$ for the right ones.

With the aid of (\ref{ham}), (\ref{bound}) the evolution operators (\ref{ev}) 
can be evaluated exactly.
In this paragraph we will suppress the Keldysh index for
simplicity. Denoting the wave function at some initial time $t_1$ as
$\psi_n(t_1,y),$  at some other time $t_2$ we find
\begin{equation}
\psi_n(t_2,y)=\psi_n(t_1,y-v_n(t_2-t_1))\chi_{nn}(t_2,t_1),
\label{psi1}
\end{equation}
for $y<0$ or $y>v_n(t_2-t_1)$ and
$$
\psi_n(t_2,y)=\sum\limits_k S_{nk}\sqrt{\frac{v_k}{v_n}}
\psi_k\left(t_1,\frac{v_k}{v_n}y-v_k(t_2-t_1)\right)
$$
\begin{equation}
\times\chi_{nk}(t_2,t_1)\chi_{nk}(t_2-y/v_n,
t_2-y/v_n), 
\label{psi2}
\end{equation}
for $0<y<v_n(t_2-t_1)$. Here we defined $ \chi_{nk}(t_2,t_1)=\exp 
(i\varphi_n(t_2)-i\varphi_k(t_1))$ and $\varphi_n(t_i)=\int_0^{t_i} dt'\, eV_n(t').$

On the other hand, by definition we have
\begin{equation}
\psi_n(t_2,y_2)=\sum\limits_k\int dy_1\; U_{nk}(t_2,t_1; y_2,y_1)\psi_k(t_1,y_1).
\label{Unk}
\end{equation}
Comparing (\ref{Unk}) with (\ref{psi1}), (\ref{psi2}) and introducing a new coordinate 
 $\tau=y/v_n$ we obtain
\begin{eqnarray}
\widehat U(t_2t_1;\tau_2\tau_1)=\delta(\tau_2-\tau_1-t_2+t_1){\rm e}^{i\hat\varphi(t_2)}
\bigg\{
\hat 1
\nonumber\\
+\,
\theta(\tau_2)\theta(-\tau_1){\rm e}^{-i\hat\varphi(t_2-\tau_2)}[\widehat S-\hat 1]{\rm e}^{i\hat\varphi(t_1-\tau_1)}
\bigg\}{\rm e}^{-i\hat\varphi(t_1)},
\label{U}
\end{eqnarray}
where $e^{i\hat \varphi }$ is the diagonal matrix with the elements $e^{i\varphi_n }$. 
Restoring the Keldysh index in (\ref{U}) we arrive at the desired result
for the evolution operators $ \widehat U_{1,2}$.

In order to evaluate the density matrix $\widehat\rho$ in the presence of interactions
one should solve a nonlinear equation (\ref{eqrho}) for arbitrary realizations of the 
fluctuating fields $V_1$ and $V_2$. In general this task cannot easily be accomplished. 
Fortunately it suffices for our present purposes to find the
density matrix for the case $V_1(t)=V_2(t)$ only. In this case
eq. (\ref{eqrho}) is trivially solved and we get
\begin{equation}
\widehat\rho(t) = \widehat U(t,0)\widehat\rho_0\widehat U(0,t),
\label{simplerho}
\end{equation}
where $\widehat U$ is defined in eq. (\ref{U}).

In order to proceed 
we will make use of the quantum Langevin equation approach \cite{Albert}.
In the case of metallic tunnel junction this approach was developed in Refs.
\cite{AES,GZ92,GZ96}. Let us define $e\dot
\varphi^+(t)=(V_{LR1}(t)+V_{LR2}(t))/2$ and $e\dot \varphi^-(t)=
V_{LR1}(t)-V_{LR2}(t)$. The key step is to treat quantum dynamics of the
$V$-fields within the quasiclassical approximation,  i.e. to assume that
fluctuations of $\varphi^-(t)$ are sufficiently small at all times.  This
assumption allows to expand the exact effective  action in powers of
$\varphi^-$ while keeping the full nonlinear dependence on the  
``center-of-mass'' field $\varphi^+$. This approximation is known 
\cite{GZ92,GZ96} to be 
particularly useful in the limit $g \gg 1$.

Expanding ${\rm Tr}\ln \widehat{\bf G}^{-1}_V$ up to the second order in $\varphi^-$ we obtain
\begin{equation}  
2{\rm Tr}\ln \widehat{\bf G}^{-1}_V \simeq 2{\rm Tr}\ln \widehat{\bf G}^{-1}|_{\varphi^-=0}+iS_R-S_I,
\label{trlog}
\end{equation}
where $iS_R={\rm Tr}\{ (\widehat G_{11}+
\widehat G_{22})\widehat{\dot\varphi}^-\}$, 
$S_I={\rm Tr}\{ \widehat G_{12} \widehat{\dot\varphi}^- 
\widehat G_{21}\widehat{\dot\varphi}^-\}$ and
$\widehat{\dot\varphi}^-$ is the diagonal matrix with the elements
$\dot\varphi^-_n$.  The zero order term in
the expansion (\ref{trlog}) vanishes.
Evaluating the first order term $iS_R$, with the aid of eqs. (\ref{Gij}),
(\ref{U}) and (\ref{simplerho}) one finds at sufficiently long times $t$
\begin{equation}
iS_R=-\frac{ig}{2\pi}\;\int\limits_0^t dt'\; \varphi^-(t')\dot\varphi^+(t').
\label{SRRR}
\end{equation}
As before, $g= 2{\rm tr}[\hat t^+\hat t]$ is the dimensionless conductance of the scatterer
and $\hat t$ is the transmission matrix. 
An analogous calculation of the second order term $S_I$ yields
\begin{eqnarray}
S_I&=&-\frac{g}{4\pi^2}\int\limits_{0}^{t}dt'\int\limits_{0}^{t}dt''
\alpha (t'-t'') \varphi^-(t')\varphi^-(t'')
\nonumber\\
&&
\times\{ \beta \cos[\varphi^+(t')-\varphi^+(t'')]
+1-\beta \},
\label{SIII}
\end{eqnarray}
where we defined $\alpha (t)=(\pi T)^2/\sinh^2[\pi Tt]$ and  
$\beta g=2{\rm tr}[\hat t^+\hat t(1-\hat t^+\hat t)]$.
Combining the results (\ref{SRRR}) and (\ref{SIII}) with the last term of eq.
(\ref{S}) we arrive at the final expression for the effective action
\begin{equation}
iS=i\int\limits_0^t dt \left[ \frac{C}{e^2}\; \dot\varphi^+\dot\varphi^- 
+\frac{I_x}{e}\;\varphi^- \right]+iS_R-S_I.
\label{finalS}
\end{equation}
In (\ref{finalS}) we also included the term which accounts for an external
current bias $I_x$.
 
{\it Quantum Langevin equation}. The action (\ref{SRRR})-(\ref{finalS})
has the same form as one for a tunnel junction with the conductance $\beta g$
shunted by an Ohmic conductor $(1-\beta )g$. Eqs. (\ref{SRRR})-(\ref{finalS})
are equivalent to the Langevin equation
\begin{equation}
\frac{C}{e}\ddot\varphi^+ + \frac{1}{eR}\dot\varphi^+-I_x= \xi_1\cos\varphi^+
+\xi_2\sin\varphi^+ +\xi_3,
\label{langevin}
\end{equation}
where the terms in the right-hand side account for the current noise and are
defined by the correlators 
\begin{equation}
\langle\xi_j(t)\xi_j(0)\rangle =
-\frac{\beta}{\pi R}\alpha (t)\left(\delta_{j1}+ \delta_{j2}+
\frac{1-\beta}{\beta}\delta_{j3}\right). 
\label{noise}
\end{equation}
In the small transparency limit eqs. (\ref{langevin}), (\ref{noise}) reduce
to those derived before for metallic tunnel junctions \cite{AES,GZ92}. If we
decompose $\varphi^+(t)=eVt+\delta \varphi^+$ ($V$ is the average voltage
across the conductor) and neglect the fluctuating part of the phase $\delta
\varphi^+$ we will immediately reproduce the well known results \cite{BB} for the
current noise in mesoscopic conductors.

{\it I-V curve}. In order to study the influence of Coulomb effects on the
current-voltage characteristics for an arbitrary scatterer we will
make use of the exact identity 
\begin{equation}
\int{\cal D}\varphi^+{\cal D}\varphi^-\;i\frac{\delta S[\varphi^+,\varphi^-]}{\delta\varphi^-(t)}
{\rm e}^{iS[\varphi^+,\varphi^-]}\equiv 0.
\label{id}
\end{equation}
Evaluating this path integral we set
$\cos[\varphi^+(t')-\varphi^+(t'')]$$=\cos[eV(t'-t'')]$ in the exponent 
of (\ref{id}) but retain the full nonlinearity in $\delta S/\delta \varphi^-$. 
This approximation works well provided either $g \gg 1$ or max$(T,eV) \gg E_C$.
A straightforward calculation then yields
\begin{equation}
I_x=\frac{V}{R}-\frac{e\beta}{\pi }\int\limits_0^{+\infty}dt\alpha (t)
{\rm e}^{-F(t)}(1-{\rm e}^{-\frac{t}{RC}})\sin[eVt],
\label{vax}
\end{equation}
\begin{eqnarray}
F(t)=-\frac{1}{g}\int_{-\infty}^{+\infty}dt'\alpha(t')
\left(\beta\cos[eVt']+1-\beta\right)
\nonumber\\
\times \left[|t'-t|-|t'|+RC\left({\rm e}^{-|t'-t|/RC}-{\rm e}^{-|t'|/RC}\right)
\right].
\label{F}
\end{eqnarray}
Eqs. (\ref{vax}), (\ref{F}) represent the central result of this paper.

{\it Single scatterer}. In the limit $g \gg 1$ and max$(eV,T) 
\gg gE_C\exp (-g/2)$ the integral in (\ref{vax}) converges at times
for which $F(t)$ is still small and can be neglected. In this limit eq. 
(\ref{vax}) yields  
\begin{equation}
I_x=\frac{V}{R}-e\beta T{\rm Im}\left[
w\Psi\left(1+\frac{w}{2} \right)-iv\Psi\left(1+\frac{iv}{2} \right)
\right].
\label{iv}
\end{equation}
where $\Psi (x)$ is the digamma function, $w=u+iv$, $u=gE_C/\pi^2T$ 
and $v=eV/\pi T$.
At $T \to 0$ from (\ref{iv}) we obtain 
\begin{equation}
R\frac{dI_x}{dV}=1-\frac{\beta}{g}\ln\left(1+\frac{1}{(eVRC)^2}
\right),
\label{dif}
\end{equation}
while in the limit $eV/E_C \gg$ max$(1,g)$ we find
\begin{equation}
RI_x=V-\beta e/2C . 
\label{offset}
\end{equation}
For $\beta =1$ the results (\ref{dif}), (\ref{offset}) reduce to
those derived in Refs. \cite{GZ92,Ark}) for tunnel junctions. Eq.
(\ref{offset}) demonstrates that at large $V$ the $I-V$ curve of
{\it any} relatively short conductor should be offset by the 
value $\beta e/2C$ due
to Coulomb effects. For instance, in disordered conductors this
offset is expected to be only 3 times smaller than for a tunnel
junction with the same $E_C$.

At $V \to 0$ from (\ref{iv}) we get 
\begin{equation}
G/G_0=1-\frac{2\beta}{g}\left[
\gamma +\Psi\left( 1+\frac{u}{2}\right)
+\frac{u}{2}\Psi'\left(1+\frac{u}{2}\right) \right], 
\label{largeg}
\end{equation}
which yields eqs. (\ref{largeT}) and (\ref{intT}) in the corresponding limits.
[The term with 1/15 in (\ref{largeT}) is recovered from (\ref{vax}),
(\ref{F}).] 

In the limit max$(eV,T) < gE_C\exp (-g/2)$ the integral (\ref{vax}) converges
at very long times and the function $F$ (\ref{F}) cannot be disregarded.
Evaluating (\ref{F}) at $t \gg 1/RC$ we find $F(t) \simeq (2/g)(\ln
(t/RC)+\gamma )$ and performing the integral in (\ref{vax}) for $g \gg 1$ we
arrive at the result (\ref{lowT})  $G/G_0=\sum_nT_n^2$.  
Hence, at extremely low $T$ the interaction correction
to $G_0$ saturates due to Coulomb and relaxation effects.  For
diffusive conductors eq. (\ref{lowT}) yields $G/G_0 \simeq 2/3$.

In the limit $g \gg 1$ our results are valid except for exponentially
low $T, eV \lesssim \tilde E_C$, in which case instanton effects
\cite{pa91,WG,Naz} gain importance and eventually turn a conductor
into an insulator at $T=0$. These effects are beyond the scope of the present
paper. For tunnel junctions the regime (\ref{lowT})  cannot be
realized since in that case $\tilde E_C/E_C \propto \exp (-g/2)$. In other
cases, however, $\tilde E_C/E_C \ll \exp (-g/2)$ and the saturation of $G_0(T)$
becomes possible. Furthermore, if the instanton effects are suppressed ($\tilde E_C
\to 0$), our results should apply down to zero temperature and voltage. 

{\it Two scatterers}. The effects discussed here can be conveniently measured
e.g. in the ``SET transistor'' configuration \cite{AL,GD} of two scatterers 
(such as, e.g., quantum point contacts) connected by a small metallic
island. With simple modifications our results hold for
such two scatterer systems as well. For instance, $G_0$ is defined
by eq. (\ref{largeg}) where $R$ is now a sum of two
resistances $R_1+R_2$, $u \to (g_1+g_2)E_C/\pi^2T$ and
$$
\frac{\beta}{g} \to \frac{\beta_1g_2+\beta_2g_1}{g_1+g_2}.
$$
The $I-V$ curve is offset at high voltages as in eq. (\ref{offset})
with $\beta \to \beta_1+\beta_2$ and $C$ being the total capacitance
of the device. Gate modulation effects can also be treated along the same lines as it was done in Ref. \cite{GZ96}. 

In summary, we studied the effect of Coulomb interaction on the $I-V$ curve of
a coherent scatterer. At low $T$ its conductance is suppressed by the universal
factor (\ref{lowT}).

We would like to thank Yu.V. Nazarov and G. Sch\"on for useful discussions.

\end{multicols}

\end{document}